\documentclass[journal]{IEEEtran}
\usepackage{color}
\usepackage{leftidx}
\usepackage{cite}
\usepackage{microtype}
\usepackage{cleveref}
\usepackage{graphicx,amssymb,amstext,amsmath,cases,subfigure,verbatim}
\usepackage[ruled,vlined,lined,ruled,linesnumbered]{algorithm2e}

\begin{document}

\title{Optimizing Throughput Fairness of Cluster-based Cooperation in Underlay Cognitive WPCNs}

\author{Lina Yuan, Suzhi Bi, Xiaohui Lin, and Hui Wang\\
\thanks{The authors are with the College of Information Engineering, Shenzhen University, Shenzhen, Guangdong, China 518060. E-mail:\{yln, bsz, xhlin, wanghsz\}@szu.edu.cn. \vspace{-2ex}}}

\maketitle

\begin{abstract}
In this paper, we consider a secondary wireless powered communication network (WPCN) underlaid to a primary point-to-point communication link. The WPCN consists of a multi-antenna hybrid access point (HAP) that transfers wireless energy to a cluster of low-power wireless devices (WDs) and receives sensing data from them. To tackle the inherent severe user unfairness problem in WPCN, we consider a cluster-based cooperation where a WD acts as the cluster head that relays the information of the other WDs. Besides, we apply energy beamforming technique to balance the dissimilar energy consumptions of the WDs to further improve the fairness. However, the use of energy beamforming and cluster-based cooperation may introduce more severe interference to the primary system than the WDs transmit independently. To guarantee the performance of primary system, we consider an interference-temperature constraint to the primary system and derive the throughput performance of each WD under the peak interference-temperature constraint. To achieve maximum throughput fairness, we jointly optimize the energy beamforming design, the transmit time allocation among the HAP and the WDs, and the transmit power allocation of each WD to maximize the minimum data rate achievable among the WDs (the max-min throughput). We show that the non-convex joint optimization problem can be transformed to a convex one and then be efficiently solved using off-the-shelf convex algorithms. Moreover, we simulate under practical network setups and show that the proposed method can effectively improve the throughput fairness of the secondary WPCN, meanwhile guaranteeing the communication quality of the primary network.
\end{abstract}


\section{Introduction}
Wireless powered communication network (WPCN) is an emerging networking paradigm that attracts extensive research attentions recently \cite{2014:Ju,2015:Bi,2017:Niyato,2016:Bi}. Specifically, the communications of wireless devices (WDs) in WPCN are powered by means of wireless power transfer \cite{2016:Bi1}, which can effectively extend the battery lifetime of low-power WDs, such as sensors and radio frequency identification tags. Compared with the traditional battery-powered communication networks, WPCN does not require manual battery replacement/charging, and thus can achieve more stable throughput performance and lower network operating costs \cite{2015:Bi}. Therefore, WPCN has found extensive applications, such as sensor networks,internet of things system, unmanned aerial vehicles communications \cite{2018:Xie}, and mobile edge computing \cite{2018:Bi}, etc. Because of the scarcity of wireless spectrum, WPCNs often need to operate in the same bandwidth with conventional wireless communication system, which may cause strong interference to each other. To tackle this problem, cognitive radio technology \cite{2009:Goldsmith} has been introduced to WPCN. Specifically, a cognitive WPCN (CWPCN) enables the WPCN (as a secondary system) to transmit opportunistically in the licensed spectrum of conventional primary wireless system \cite{2013:Lee3} \cite{2014:Hoang}. For efficient utilization of spectrum resources, Lee \cite{2015:Lee1} considered a secondary WPCN both underlaid and overlaid with a primary system. In particular, in an underlay case, the secondary system is not aware of the channel state information (CSI) of the primary system, thus it can only restrict the interference power generated to the primary system, i.e., the interference temperature constraint (ITC). In comparison, in an overlay case, the secondary system knows the primary system's CSI, e.g., via primary system's feedback, so it can control its own transmission to meet the data rate requirement of the primary system, i.e., the primary rate constraint. Under the ITC, Cheng \cite{2017:Cheng} considered a CWPCN that uses a harvest-then-transmit protocol based on time division multiple access for secondary system and proposed an algorithm that optimizes the secondary system user throughput fairness. Yin \cite{2017:Yin} designed a cooperative spectrum sharing model of a CWPCN, which exploits the cooperation between primary and secondary systems to maximize the energy efficiency of secondary users. Xu \cite{2017:Xu} investigated cooperative resource allocation in a multi-carrier CWPCN, which the primary user harvests energy from the information signals transmitted by the secondary users.

In WPCN, it is common to use hybrid access point (HAP) that transfers wireless energy to a cluster of low-power WDs and receives wireless information from them. It is well-known that a severe ``doubly-near-far" user unfairness problem exists in WPCN that the data rates of users that are farther away from HAP is significantly lower than that of the adjacent users \cite{2014:Ju}. User cooperation is a commonly used method to enhance the user fairness issue, which in fact can potentially benefit all the participating users. Intuitively, the remote users can improve the link quality to the HAP with the help from near users. While for near users, the data rate loss caused by helping the others can be made up from the longer energy transfer time allocated by the HAP, because now the far users need less time to harvest energy when cooperation is applied. A number of practical user cooperation has been proposed for WPCNs \cite{2017:Liang,2014:Ju1,2015:Chen,2017:Zhong,2017:Yuan}. For instance, a two-user cooperation WPCN was firstly investigated by Ju \cite{2014:Ju1}, where the authors showed that close-to-HAP user can improve its data rate by helping the far user, achieving a win-win situation. Subsequently, Zhong \cite {2017:Zhong} considered a pair of distributed end users first harvesting energy from an energy node, and then transmit jointly their information to a destination node using distributed space-time code. Further, Chen \cite{2015:Chen} first presented a harvest-then-cooperate protocol in WPCNs for a three-node reference model, and latter extended to a general multiple user cooperation scenario. In addition, Yuan \cite{2017:Yuan} proposed a multi-antenna enabled cluster-based cooperation, where one WD acts as the cluster head (CH), e.g., selecting the one nearest to the cluster center, to relay the information transmission of the other cluster members (CMs, the remaining ones except the CH) to the HAP. Meanwhile, energy beamforming (EB) technique \cite{2014:Liu} is used at the multi-antenna HAP to achieve directional energy transfer for balancing the energy consumptions of different WDs. It showed that the cluster-based cooperation can effectively enhance both the user fairness and spectral efficiency of WPCN, compared to some other representative benchmark cooperation methods. Therefore, we adopt the cluster-based cooperative WPCN in this paper. However, the use of cooperation and EB, although effective in enhancing the throughput of WPCN, may also cause strong interference to the primary link. The achievable throughput performance of cluster-based cooperation is not known when interference is taken into consideration.

As shown in Fig. 1, we consider in this paper a WPCN underlaid to a primary communication link. The WDs in the WPCN use a cluster-based cooperation to transmit to the HAP. The detailed contributions of this paper are as follows.
\begin{itemize}
  \item We consider a time division multiple access based protocol that firstly the HAP broadcasts energy to a cluster of WDs and then the WDs transmit their information back to the HAP using the harvested energy. Here, one of the WDs is selected as the CH that relays the messages of the other WDs. EB is applied at the HAP for both enhancing the energy transfer efficiency in the downlink and the spectral efficiency in the uplink information transmission. We derive the throughput of each individual WD under the interference from the primary communication link. Meanwhile, to control the interference to the primary link in the underlay scenario, we consider a peak ITC such that the interference generated by the WPCN is limited to a prescribed threshold.
  \item To improve the user fairness among the WDs, we jointly optimize the beamforming of the HAP, the transmit time allocation among the HAP and the WDs, and the transmit power allocation of the WDs to maximize the minimum data rate achievable among the overall WDs (the max-min throughput) under the peak ITC. We show that the non-convex joint optimization problem can be transformed to a convex one and subsequently solved using off-the-shelf convex algorithms.
  \item We simulate under practical network setups and show that the proposed method can significantly enhance the throughput fairness of the WPCN under the peak ITC compared to the representative benchmark method. In particular, the advantageous performance is most evident under stringent ITC requirement. Intuitively, this is because the cluster-based cooperation can effectively control its interference to the primary system by reducing the communication range, and thus the transmit power, when the secondary users transmit their information.
\end{itemize}

The rest of the paper is organized as follows. In Section II, we introduce the system model and the cluster-based cooperation method. We analyze the secondary throughput performance in Section III. In Section IV, we formulate the maxi-min throughput optimization problem and transform the non-convex problem into the convex one. In Section V, simulation results are provided to evaluate the performance of the proposed cooperation, followed by concluding remarks in Section VI.
\begin{figure}
  \centering
  \includegraphics[width=3.4in]{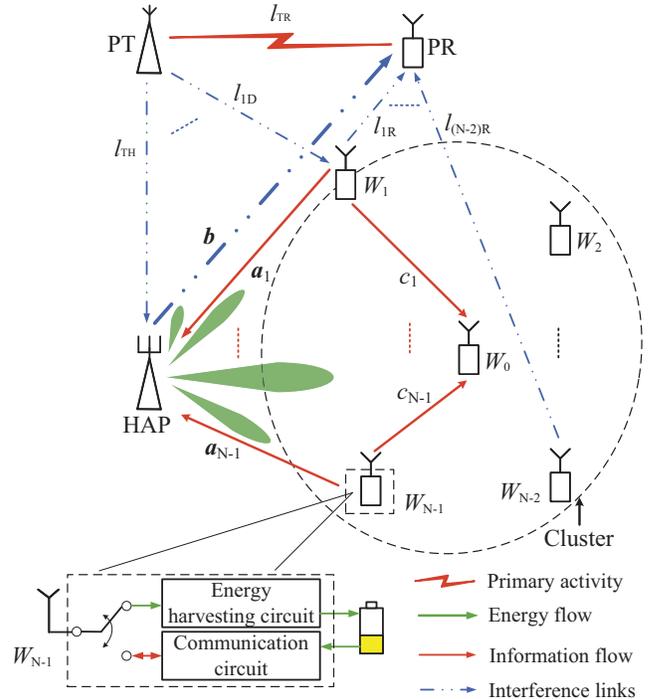}\\
  \caption{A schematic of the considered cluster-based cooperation in an underlay CWPCN, where W$_0$ is the CH and the rest $(N-1)$ WDs are CMs.}
  \label{101}
\end{figure}

\section{System Model}
\subsection{Channel Model}
As shown in Fig.1, we consider a CWPCN coexisting with a primary link consisting of a pair of single-antenna primary transmitter (PT) and primary receiver (PR). The CWPCN consists of one HAP and $N$ WDs. The HAP is assumed to be equipped with $M$ antennas and the WDs each has a single antenna. The HAP can implement EB in the downlink for directional energy transfer and MRC in the uplink to enhance spectrum efficiency \cite{2014:Liu}. Specifically, the HAP broadcasts wireless energy and receives wireless information transmission (WIT) to/from the WDs. We assume that both the primary and secondary systems operate over the same frequency band. For the secondary system, a time division duplexing circuit is implemented at both the HAP and the WDs to separate the energy and information transmissions. Meanwhile, the primary communication link is active throughout the considered time interval.

In this paper, one of the WDs is selected as the CH that helps relay the WIT of the other CMs, e.g., selecting the WD closest to the cluster center. The impact of CH selection method will be discussed in simulations. Without loss of generality, the CH is denoted by W$_0$, and the CMs are denoted by W$_1$, $\cdots$, W$_{N-1}$. All the channels are assumed to be independent and reciprocal and follow quasi-static flat-fading, such that all the channels coefficients remain constant during each block transmission time, denoted by $T$, but can vary in different blocks. The channel coefficient vector between the HAP and W$_{i}$ is denoted by $\mathbf{a}_{i} \in\mathcal{C}^{M\times1}$,  where $\mathbf{a}_{i} \sim \mathcal{CN} (\mathbf{0}, \sigma_{i}^2 \mathbf{I})$ and $\sigma_i^2$ denotes the average channel gain, $i=0,1,\cdots, N-1$. Besides, the channel coefficients between the $j$-th CM and the CH are denoted by $c_j\sim \mathcal{CN} ( 0, \delta_{j}^2)$, $j=1,\cdots, N-1$. Meanwhile, the channel coefficients, between the PT and PR, and between the PT and HAP, are denoted by $l_{TR} \sim \mathcal{CN} ( 0, \delta_{TR}^2)$ and $l_{TH} \sim \mathcal{CN} ( 0, \delta_{TH}^2)$, respectively. Denote $\mathbf{b} \in\mathcal{C}^{M\times1}$ as the interference channel coefficients between the HAP and PR, where $\mathbf{b} \sim \mathcal{CN} (\mathbf{0}, \sigma^2 \mathbf{I})$. Let $l_{iR}\sim \mathcal{CN} (0, \delta_{iR}^2)$ and $l_{iD}\sim \mathcal{CN} (0, \delta_{iD}^2)$, $i=0,1,\cdots, N-1$, represent the interference channel coefficient vector between the PR and WDs, and between the PT and WDs, respectively. Here, we use $h_{i} \triangleq |\mathbf{a}_{i}|^2$, $g_i\triangleq |c_i|^2$, $h_{iR}\triangleq |l_{iR}|^2$, $h_{iD}\triangleq |l_{iD}|^2$, $h_{TR}\triangleq |l_{TR}|^2$, and $h_{TH}\triangleq |l_{TH}|^2$ to denote the corresponding channel gains, where $|\cdot|$ denotes the 2-norm operator.
\begin{figure}
  \centering
  \includegraphics[width=3.5in]{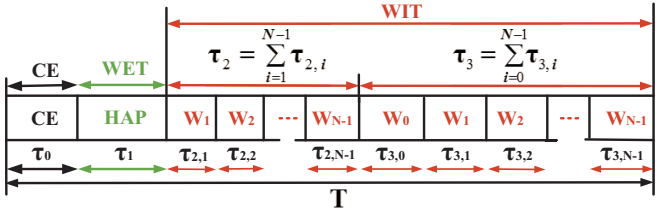}\\
  \caption{The proposed cluster-based cooperation protocol in U-CWPCN.}
  \label{102}
\end{figure}

\subsection{Cluster-based Cooperation Protocol}
The operation of the proposed cluster-based cooperation in a transmission time block is illustrated in Fig.~\ref{102}. At the beginning of a transmission block, channel estimation (CE) is performed within a fixed duration $\tau_{0}$. Through CE, we assume that the HAP has the knowledge of CSI inside the WPCN, i.e., $\mathbf{a}_i$'s, and $c_i$'s, which are acquired from pilot transmissions and CSI feedback from the WDs. Besides, we assume that the operating protocol of the primary system (including the transmit power of the PT and PR) is known, e.g., through a separate control channel. Therefore, the interference channels between the primary system and the secondary WPCN can be estimated by the HAP and WDs (like in \cite{2017:Cheng}), e.g., by estimating the pilot signals sent by the PT and PR. However, the channel between the PT and PR is private information in our underlay framework and unknown by the HAP.

After the CE stage, the HAP coordinates the secondary WPCN to operate in three phases. In the first phase with time duration $\tau_{1}$, the HAP broadcasts wireless energy to the WDs. In the second phase, the $N-1$ CMs transmit its own information in turn to the CH, where the $i$-th CM transmits for $\tau_{2,i}$ amount of time, $i=1, \cdots, N-1$. In the third phase, the CH transmits the decoded messages of the $N-1$ CMs along with its own message to the HAP, where the time taken to transmit the $i$-th WD's message is denoted as $\tau_{3,i}$, $i=0,1,\cdots, N-1$. Evidently, the time allocations during a transmission block satisfy the following inequality
\begin{equation}
\label{1}
    \tau_0+\tau_1+\sum^{N-1}_{i=1}\tau_{2,i}+\sum^{N-1}_{i=0}\tau_{3,i} \leq T.\\
\end{equation}

Notice that $\tau_0$ is a known parameter. Without loss of generality, we assume $T=1$ throughout this paper.

\subsection{Cognitive Underlay Transmissions}
All WDs are assumed to have no fixed power supplies, and thus need to replenish energy by harvesting RF energy from the HAP's wireless energy transmission in the downlink; the harvested energy at each WD is stored in a rechargeable battery and then used for its WIT in the uplink. The PT and the HAP are assumed to have stable power supplies. We assume that the PT transmits with constant power $P_p$, and the HAP's maximum transmit power is $P_{max}$.

With the CSI knowledge at the HAP, it jointly optimizes the system resource allocation within the WPCN, e.g., transmission durations, power, and the beamforming design, and coordinate the transmissions of all the WDs. Specifically, we can maximize the throughput of the secondary WPCN, meanwhile guaranteeing that the resulting interference generated to the primary link is below a predefined threshold. In the following, we formulate the throughput maximization problem and propose an efficient method to solve it optimally.

\section{Secondary Throughput Analysis}
In this section, we derive the throughput of each secondary WD achieved by the proposed cluster-based cooperation protocol. The results will be used in the next section to optimize the throughput fairness of the U-CWPCN.

\subsection{Phase I: Energy Transfer}
During the WET phase, the HAP adopts EB to deliver different levels of wireless power to distributed WDs to balance their different energy consumptions. Specifically, in the first phase of time $\tau_1$, the HAP transmits $\mathbf{w}(t) \in C^{M\times 1}$ random energy signals on the $M$ antennas, where the transmit power of HAP is constrained by $P_H$ as
\begin{equation}
\label{2}
E\left[|\mathbf{w}(t)|^2\right] = \text{tr}\left(E\left\{\mathbf{w}(t)\mathbf{w}(t)^H\right\}\right) \triangleq \text{tr}(\mathbf{Q}) \leq P_H,
\end{equation}
where $\text{tr}(\cdot)$ denotes the trace of a square matrix, $(\cdot)^H$ denotes the complex conjugate operator, and $\mathbf{Q}\succeq \mathbf{0}$ is the beamforming matrix.

Then, the received energy by the $i$-th WD is \cite{2015:Lee1}
\begin{equation}
\label{3}
H_{i}= \eta \tau_1 \cdot \text{tr}(\mathbf{A}_i\mathbf{Q}),
\end{equation}
here, $\mathbf{A}_i \triangleq \mathbf{a}_i\mathbf{a}_i^H$ and $\eta\in(0,1]$ denotes the energy harvesting efficiency, which is assumed equal for all the WDs. Accordingly, the residual energy of the $i$-th WD is

\begin{equation}
\label{4}
E_{i} = \min\left\{E_{0,i} + H_i, E_{max}\right\},
\end{equation}
where, $E_{0,i}$ is the known residual energy at the beginning of the current time slot, and $E_{max}$ is the battery capacity.

In the meantime, the interference power to the primary receiver is $\text{tr}(\mathbf{bb}^H \mathbf{Q})\triangleq \text{tr}(\mathbf{H}_{HR} \mathbf{Q})$, where $\mathbf{H}_{HR} \triangleq \mathbf{bb}^H$.

\subsection{Phase II: Intra-cluster Transmissions}
In general, the CMs only use a part of the harvested energy to transmit to the CH. The transmit power of the $i$-th CM $P_{2, i}$ is restricted by
\begin{equation}
\label{5}
\tau_{2,i} P_{2, i} + e_i \leq E_{i},\  i=1,\cdots,N-1,
\end{equation}
where $e_i$ denotes the fixed circuit energy consumption of the $i$-th CM within a transmission block, e.g., on performing sensing and data processing.
The received signal at the CH is expressed as
\begin{equation}
\label{6}
\begin{aligned}
   y_{0,i}^{(2)}(t)= c_i \sqrt{P_{2, i}}s_{i}^{(2)}(t)+ n_{i}^{(2)}(t),
\end{aligned}
\end{equation}
where $n_{i}^{(2)}(t)$ includes the receiver noise $N_0$ and the interference signal generated from the primary link $h_{0D}P_p$. The total interference plus noise power $E\left[|n_{i}^{(2)}(t)|^2\right]=N_0+h_{0D}P_p$. In this paper, we consider the worst-case interference in terms of the achievable data rate given the noise power, where the interference distribution is cyclic symmetric complex Gaussian \cite{2006:Cover}. Then, the CH can decode the $i$-th CM's message at the minimum achievable rate given by \footnote{For simplicity of illustration in the following analysis, we use the term achievable data rate to represent the minimum achievable data rate by treating the interference signal as cyclic symmetric complex Gaussian.}
\begin{equation}
\label{7}
\begin{aligned}
R_{i}^{(2)}=\tau_{2,i} \log_{2}\left(1 + \frac{g_i P_{2,i}}{N_0 + h_{0D}P_p}\right),i=1,\cdots,N-1.
\end{aligned}
\end{equation}

Meanwhile, the HAP can also overhear the transmission of the CMs, such that it receives
\begin{equation}
\label{8}
\mathbf{y}_{H,i}^{(2)}(t) = \mathbf{a}_i \sqrt{P_{2, i}}s_{i}^{(2)}(t) + \mathbf{n}_{H,i}^{(2)}(t),
\end{equation}
during the $i$-th CM's transmission, where $i=1,\cdots,N-1$, and $\mathbf{n}_{H,i}^{(2)}(t) \sim \mathcal{CN}\left(\mathbf{0},(N_0+h_{TH}P_p)\mathbf{I}\right)$, $h_{TH} P_p$ is the interference power to the HAP from the primary link.

Meanwhile, the interference power caused by the $i$-th WD to the primary system is $h_{iR} P_{2,i}$, $i=1,\cdots, N-1$.

\subsection{Phase III: Cluster-to-HAP Transmission}
After decoding the CMs' messages, the CH transmits the $(N-1)$ CMs' messages along with its own message one by one to the HAP. Let $P_{3,i}$ denote the power used to transmit the $i$-th WD's message. Then, the received signal of the $i$-th WD's message at the HAP is
\begin{equation}
\label{9}
\mathbf{y}_{i}^{(3)}(t) = \mathbf{a}_0 \sqrt{P_{3,i}}s_{i}^{(3)}(t) + \mathbf{n}_{i}^{(3)}(t),
\end{equation}
where $\mathbf{n}_{i}^{(3)}(t) \sim \mathcal{CN}\left(\mathbf{0},(N_0+h_{TH}P_p)\mathbf{I}\right)$, and $i=0,1,\cdots, N-1$. The total energy consumed by CH is upper bounded by its harvested energy
\begin{equation}
\label{10}
\sum^{N-1}_{i=0}\tau_{3,i} P_{3,i} + e_0 \leq E_{0},
\end{equation}
where $e_0$ is the total energy consumed by the other modules of the CH within a transmission block, such as the central processing unit, and the passive power on circuitry.

We assume that the HAP uses MRC to maximize the received signal to interference plus noise power ratio (SINR), where the combiner output SINR of the $i$-th WD is
\begin{equation}
\label{11}
\begin{aligned}
\gamma^{(3)}_{i} = \frac{|\mathbf{a}_0|^2 P_{3,i}}{N_0 + h_{TH} P_p}=\frac{h_0 P_{3,i}}{N_0 + h_{TH} P_p}, \ i=0,\cdots, N-1,
\end{aligned}
\end{equation}
where the numerator denotes the useful signal power for the $i$-th user, and the denominator is the sum of the interference power generated by the primary link and the receiver noise power. At the same time, the interference from the CH to the PR is $h_{0R} P_{3,i}$, $i=0,\cdots, N-1$, owing to CH's WIT.

We denote the time allocation as $\pmb{\tau}=[\tau_1, \tau_{2,1}, \cdots, \tau_{2,N-1}, \tau_{3,0}, \tau_{3,1}, \cdots, \tau_{3,N-1}]'$, and the transmit power as $\pmb{P}=[P_{2,1}, \cdots, P_{2,N-1}, P_{3,0}, P_{3,1}, \cdots, P_{3,N-1}]'$, where $[\cdot]'$ denotes the transpose operator. Then, the data rate of the CH at the HAP is
\begin{equation}
\label{12}
\begin{aligned}
R_{0}(\pmb{\tau}, \pmb{P})= \tau_{3,0} \log_{2}\left(1 + \frac{h_0 P_{3,0}}{N_0 + h_{TH} P_p}\right).
\end{aligned}
\end{equation}

For each CM's message, however, is received in both the second and third phases. In this case, the HAP can jointly decode each CM's message across two phases at a rate given by \cite{2014:Ju}
\begin{equation}
\label{13}
R_{i}(\pmb{\tau}, \pmb{P}) = \min\left\{R_{i}^{(2)}(\pmb{\tau}, \pmb{P}), V_{i}^{(2)}(\pmb{\tau}, \pmb{P}) + V_{i}^{(3)}(\pmb{\tau}, \pmb{P})\right\},
\end{equation}
where $i=1, \cdots,N-1$, and $R_{i}^{(2)}(\pmb{\tau}, \pmb{P})$ is given in (\ref{7}).
$V_{i}^{(2)}(\pmb{\tau}, \pmb{P})$ denotes the information that can be extracted by the HAP from the received signal in (\ref{6}) (in the second phase) using an optimal MRC receiver, which is given by
\begin{equation}
\label{14}
\begin{aligned}
V_{i}^{(2)}(\pmb{\tau}, \pmb{P})=\tau_{2,i} \log_{2}\left(1 + \frac{ h_i P_{2,i}}{N_0 +  h_{TH}P_p}\right),
\end{aligned}
\end{equation}
$V_{i}^{(3)}(\pmb{\tau}, \pmb{P})$ denotes the achievable rates of the transmissions from CH to the HAP, which are given by
\begin{equation}
\label{15}
\begin{aligned}
V_{i}^{(3)}(\pmb{\tau}, \pmb{P})=\tau_{3,i} \log_{2}\left(1+\frac{h_0 P_{3,i}}{N_0 + h_{TH} P_p}\right).
\end{aligned}
\end{equation}

An important performance metric of an U-CWPCN is the minimum achievable secondary throughput among all the WDs (the max-min throughput), i.e.,
\begin{equation}
\label{16}
S = \min_{0 \leq i \leq N-1} R_i,
\end{equation}
which reflects the throughput fairness. The max-min throughput has important practical implication. For instance, the max-min throughput in a wireless sensor network reflects the accuracy of data reported by the ``bottleneck" sensor, which can directly affect the overall sensing accuracy of the network. In the next section, we formulate the max-min throughput optimization problem and solve it optimally. In fact, our proposed method in this paper can also be extended to maximize (weighted) sum throughput of the WDs, which is omitted for brevity.

\section{Max-min Throughput Optimization Under the peak interference constraint}
\subsection{Problem Formulation}
In this subsection, we require that the interference power from the WPCN to the PR is no larger than a predefined threshold, denoted by $I_{max}$. Therefore, the interference constraints of the three phases are, respectively,
\begin{equation}
\label{17}
\text{tr}(\mathbf{h}_{HR} \mathbf{Q}) \leq I_{max}.
\end{equation}
\begin{equation}
\label{18}
I^{CM}_i= h_{iD} P_{2,i} \leq I_{max}, \ i=1,\cdots, N-1.
\end{equation}
\begin{equation}
\label{19}
I^{CH}_i=h_{0D} P_{3,i} \leq I_{max}, \ i=0,\cdots, N-1.
\end{equation}

Under this setup, we are interested in maximizing the minimum (max-min) throughput of all WDs in each block, by jointly optimizing the EB $\mathbf{Q}$, the time allocation $\pmb {\tau}$, and the transmit power allocation $\pmb P$, i.e.,
\begin{equation}
\label{20}
   \begin{aligned}
(P1):\; &\max_{\pmb {\tau, P}, \mathbf{Q}}& &S= \min_{0 \leq i \leq N-1} R_i(\pmb {\tau, P})\\
&\text{\;\;s. t.}& & (1), (2),(4),(5),(10),(17),(18), \rm{and}\; (19), \\
                 & & & \tau_{1} \geq 0, \; \tau_{2,i} \geq 0, \;i=1,\cdots, N-1,\\
                 & & & \tau_{3,i} \geq 0,\;P_{3,i}\geq 0,\;i=0,1,\cdots, N-1,\\
                 & & & \mathbf{Q} \succeq \mathbf{0}.\\
    \end{aligned}
\end{equation}

By introducing an auxiliary variable $\overline S$, problem (\ref{20}) can be equivalently transformed into its epigraphic form,
\begin{equation}
\small
\label{21}
   \begin{aligned}
(P2):\; &\max_{\pmb {\tau}, \pmb{P}, \mathbf{Q}, \overline S} & &  \overline S\\
 &\text{\;\;\;\;s. t.}  & & (1), (2),(4),(5),(10),(17),(18), \rm{and}\; (19), \\
                   & & &\pmb {\tau} \geq \mathbf{0}, \; {\pmb {P} \geq \mathbf{0}},\; \mathbf{Q} \succeq \mathbf{0}, \;\overline S \geq 0,\\
                   & & & R_{0}(\pmb{\tau}, \pmb{P}) \geq \overline S,\\
                   & & &V_{i}^{(2)}(\pmb{\tau}, \pmb{P}) + V_{i}^{(3)}(\pmb{\tau}, \pmb{P}) \geq \overline S,\\
                   & & &R_{i}^{(2)}(\pmb{\tau}, \pmb{P}) \geq \overline S, i=1,\cdots, N-1.\\
    \end{aligned}
\end{equation}

Notice that both the data rates expressions of intra-cluster communication (i.e., $R_{i}^{(2)}(\pmb{\tau}, \pmb{P})$ and $V_{i}^{(2)}(\pmb{\tau}, \pmb{P}))$ and cluster-to-HAP communication (i.e., $R_{0}(\pmb{\tau}, \pmb{P})$ and $V_{i}^{(3)}(\pmb{\tau}, \pmb{P})$) are not concave functions. Besides, the LHS of (\ref{5}) is also not a convex function. Therefore, (P2) is a non-convex problem in its current form, which lacks of efficient optimal algorithms. In the next subsection, we transform the above non-convex problem into a convex one through introducing some auxiliary variables, which can then be solved using some known convex optimization techniques, e.g., interior point method \cite{2004:Boyd}.
\subsection{Convex Transformation of (P2)}
The basic idea of the convex transformation is to introduce auxiliary variables to replace the multiplicative terms in (5) and (10). Specifically, we first define $\Psi_{2,i} \triangleq \frac{\tau_{2,i}P_{2,i}}{\eta}$, $i=1,\cdots, N-1$ and $\pmb {\Psi}=\left[\Psi_{2,1}, \cdots, \Psi_{2,N-1}\right]'$. Accordingly, (\ref{5}) can be re-written as a function of $\pmb {\Psi}$,
\begin{equation}
\small
\label{22}
{\Psi}_{2,i} + \frac{e_{i}}{\eta} \leq \frac{E_i}{\eta}, i=1,\cdots, N-1.
\end{equation}

Meanwhile, $R_{i}^{(2)}(\pmb{\tau}, \pmb{P})$ and $V_{i}^{(2)}(\pmb{\tau}, \pmb{P})$ in (\ref{7}) and (\ref{14}) can be re-expressed as functions of $\pmb {\tau}$ and $\pmb {\Psi}$, respectively,
\begin{equation}
\label{23}
R_{i}^{(2)}(\pmb{\tau},\pmb {\Psi})=\tau_{2,i} \log_{2}\left(1+ \overline{\rho}_i \frac {{\Psi}_{2,i}}{\tau_{2,i}}\right),
\end{equation}
\begin{equation}
\small
\label{24}
V_{i}^{(2)}(\pmb{\tau},\pmb {\Psi})=\tau_{2,i} \log_{2}\left(1+ \rho_i \frac {{\Psi}_{2,i}}{\tau_{2,i}}\right),
\end{equation}
where $\overline {\rho}_i\triangleq\eta \frac{g_i}{N_0 + h_{0D}P_p}$, $\rho_i\triangleq\eta \frac{h_i}{N_0 + h_{TH}P_p}$ are parameters, and $i=1,\cdots,N-1$.

Subsequently, we define $\theta_{3,i}\triangleq\frac{\tau_{3,i} P_{3,i}}{\eta}$, $i=0,1,\cdots,N-1$, and $\pmb {\theta}=\left[\theta_{3,0}, \cdots, \theta_{3,N-1}\right]'$, then $R_{0}(\pmb{\tau}, \pmb{P})$ and $V_{i}^{(3)}(\pmb{\tau}, \pmb{P})$ in (\ref{12}) and (\ref{15}) can be reformulated as  functions of $\pmb {\tau}$ and $\pmb {\theta}$, i.e.,
\begin{equation}
\label{25}
\begin{aligned}
R_{0}(\pmb {\tau, \theta})= \tau_{3,0} \log_{2}\left(1 + \rho_0 \frac { \theta_{3,0}}{\tau_{3,0}}\right),
\end{aligned}
\end{equation}
\begin{equation}
\label{26}
\begin{aligned}
V_{i}^{(3)}(\pmb {\tau, \theta})=\tau_{3,i} \log_{2}\left(1+\rho_0 \frac { \theta_{3,i}}{\tau_{3,i}}\right),
\end{aligned}
\end{equation}
where $i=1,\cdots,N-1$, and $\rho_0\triangleq\eta \frac{h_0}{N_0 + h_{TH} P_p}$.

Next, we define $\mathbf{W} \triangleq \tau_1\mathbf{Q} \succeq \mathbf{0}$. With the sum transmit power constraint in (\ref{3}), we have
\begin{equation}
\label{27}
\text{tr}\left(\mathbf{W}\right) = \text{tr}\left(\tau_1\mathbf{Q}\right) \leq \tau_1 P_H.
\end{equation}
Accordingly, we change the variables as
\begin{equation}
\label{28}
z_i \triangleq \tau_1\text{tr}\left(\mathbf{A}_i\mathbf{Q}\right) = \text{tr}\left(\mathbf{A}_i\mathbf{W}\right),
\end{equation}
for $i=0,\cdots, N-1$. Thus, the power constraint given in (\ref{10}) can be re-expressed as
\begin{equation}
\label{29}
\begin{aligned}
\sum^{N-1}_{i=0}\theta_{3,i} + \frac{e_{0}}{\eta} \leq z_0.
\end{aligned}
\end{equation}
At the same time, (\ref{17}), (\ref{18}), and (\ref{19}) can be reformed as, respectively,
\begin{equation}
\label{30}
\text{tr}(\mathbf{h}_{HR} \mathbf{W}) \leq \tau_1 I_{max},
\end{equation}
\begin{equation}
\label{31}
\begin{aligned}
\Psi_{2,i} \leq \tau_{2,i} \frac{I_{max}}{\overline{\phi}_i}, i=1,\cdots,N-1,
\end{aligned}
\end{equation}
\begin{equation}
\label{32}
\begin{aligned}
\theta_{3,i} \leq \tau_{3,i} \frac{I_{max}}{\phi_0},i=0,1,\cdots,N-1,
\end{aligned}
\end{equation}
where $\phi_0 = \eta h_{0D}$, and $\overline{\phi}_i = \eta h_{iD}$, $i=1,\cdots,N-1$.

Accordingly, problem (\ref{21}) can be transformed into the following equivalent problem.
\begin{equation*}
\label{33}
   \begin{aligned}
(P3): \\
\max_{\pmb {\tau}, \pmb{\theta}, \pmb {\Psi}, \pmb z, \overline S, \mathbf{W}\succeq \mathbf{0}} &  \overline S\\
\text{s. t.\quad \quad \;}   & (22), (28),(29),(30),(31),\rm{and}\;(32), \\
                   & \pmb {\tau} \geq \mathbf{0}, \; \pmb{\theta} \geq 0, \; \overline S \geq 0,\;\mathbf{W} \succeq \mathbf{0},\;\text{tr}(\mathbf{W}) \leq \tau_1 P_H,\\
                   & R_{0}(\pmb{\tau, \theta}) \geq \overline S,\;V_{i}^{(2)}(\pmb{\tau},\pmb {\Psi}) + V_{i}^{(3)}(\pmb{\tau, \theta}) \geq \overline S,\\
                   & R_{i}^{(2)}(\pmb{\tau},\pmb{\Psi}) \geq \overline S,\ i=1,\cdots, N-1,\\
                   & \tau_0+\tau_1+\sum^{N-1}_{i=1}\tau_{2,i}+\sum^{N-1}_{i=0}\tau_{3,i} \leq 1,\\
                   & E_i \leq E_{0,i} + \eta z_i, \ i=0,1,\cdots,N-1,\\
                   & E_i \leq E_{max}, \ i=0,1,\cdots,N-1.\\
    \end{aligned}
\end{equation*}

Furthermore, our proposed method in this paper can also be extended to be under the average ITC, which is omitted for brevity. From \cite{2017:Yuan}, both $R^{(2)}_{i}$'s in (\ref{23}) and $V_{i}^{(2)}$'s in (\ref{24}) are concave functions in $(\pmb{\tau},\pmb{\Psi})'$. Besides, $R_{0}$ in (\ref{25}) and $V_{i}^{(3)}$'s in (\ref{26}) are also concave functions in $(\pmb {\pmb{\tau}, \theta})'$. Therefore, the first three sets of constraints in (P3) are convex constraints. Meanwhile, the rest of the constraints are affine. It follows that the objective and all the constraints of (P3) are convex, therefore (P3) is a convex optimization problem, which can be efficiently solved by off-the-shelf optimization algorithms, e.g., interior point method \cite{2004:Boyd}. Let's denote the optimal solution to (P3) as $\left\{\pmb {\tau}^*, \pmb{\theta}^*, \pmb {\Psi}^*, \pmb z^*, \overline S^*, \mathbf{W}^*\right\}$. Then, the optimal solution $\pmb {\tau}^*$ of (P1) is the same as that in (P3). The optimal $\mathbf{Q}^*$ and $\pmb{P}^*$ of (P1) can be restored by letting $\mathbf{Q}^* = \mathbf{W}^*/\tau_1^*$, $P_{2,i} = \eta \pmb {\Psi}^* / \tau_{2,i}^*$ ($\ i = 1,\cdots,N-1$), and $P_{3,i}^* = \eta\theta_{3,i}^*/\tau_{3,i}^*$ ($\ i = 0,\cdots,N-1$).

\begin{figure}
  \centering
  \includegraphics[width=3.4in]{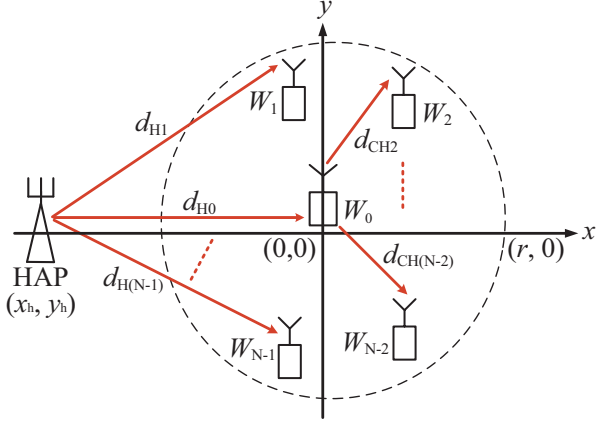}\\
  \caption{The deployment of HAP and WDs in C-WPCN, where $N=15$.}
  \label{106}
\end{figure}
\subsection{Benchmark Methods}
For performance comparison, we consider two representative benchmark methods, i.e., independent transmission and hybrid transmission. For simplicity, we assume that the time spent on CE ($\tau_0$) is equal for all the schemes.
\subsubsection{Independent transmission}
In this case, all the WDs transmit independently to the HAP following the harvest-then-transmit protocol in \cite{2015:Chen}. Specifically, the HAP first uses EB to perform WET for $\tau_1'$ amount of time for the WDs to harvest. Then, the WDs take turns to transmit their messages to the HAP, where each WD's transmission takes $\tau_{2,i}'$ ($i'=0,1,\cdots, N-1$) amount of time. Meanwhile, the HAP uses MRC to decode the message of each user.\footnote{Spatial multiplexing is not used at the HAP as the number of WDs is often much larger than the number of antennas at the HAP. Otherwise, either strong interference or high computational complexity will be induced when the WDs transmit to the HAP simultaneously.} Then, the data rate of the $i$-th user is denoted by
\begin{equation}
\label{34}
R_{i}'(\pmb{\tau}', \mathbf{P'}) = \tau_{2,i}'  \log_{2}\left(1 + \frac{h_i P_{2,i}'}{N_0 + h_{TH}P_p}\right), \
\end{equation}
where $i=0,\cdots,N-1$, $\pmb{\tau}' \triangleq[\tau_1', \tau_{2,0}', \cdots, \tau_{2,N-1}']'$, and $\pmb{P}' \triangleq[P_{2,0}', \cdots, P_{2,N-1}']'$. Then, the max-min throughput can be obtained by solving the following problem
\begin{equation}
\small
\label{36}
   \begin{aligned}
 &\max_{\mathbf{\pmb {\tau}'}, \pmb{P}', \mathbf{Q}'}& &\min_{i=0,\cdots, N-1} R_{i}'(\pmb{\tau}', \pmb{P}')\\
        &\text{s. t.}& & \tau_0'+\tau_1'+\sum^{N-1}_{i=0}\tau_{2,i}' \leq 1, \\
                 & & & \tau_{1}' \geq 0,\;\tau_{2,i}'\geq 0,\;\mathbf{Q}' \succeq \mathbf{0},\;\text{tr}(\mathbf{Q'}) \leq P_H,\\
                 & & & \text{tr}(\mathbf{h}_{HR} \mathbf{Q'}) \leq I_{max}, \; h_{iD} P_{2,i}' \leq I_{max}, \\
                 & & & \tau_{2,i}' P_{2,i}' + e_i \leq E_i', \; E_i' \leq E_{max}, \\
                 & & & E_i' \leq E_{0,i}' + \eta \tau_1'\text{tr}(\mathbf{A}_i \mathbf{Q'}),\; i=0,\cdots, N-1.\\
    \end{aligned}
\end{equation}
The optimal solution to the above problem can be similarly obtained as (P3), where the detailed algorithm is omitted for simplicity.

\begin{figure}
\centering
\subfigure[]{
\label{figa} 
\includegraphics[width=3.4in]{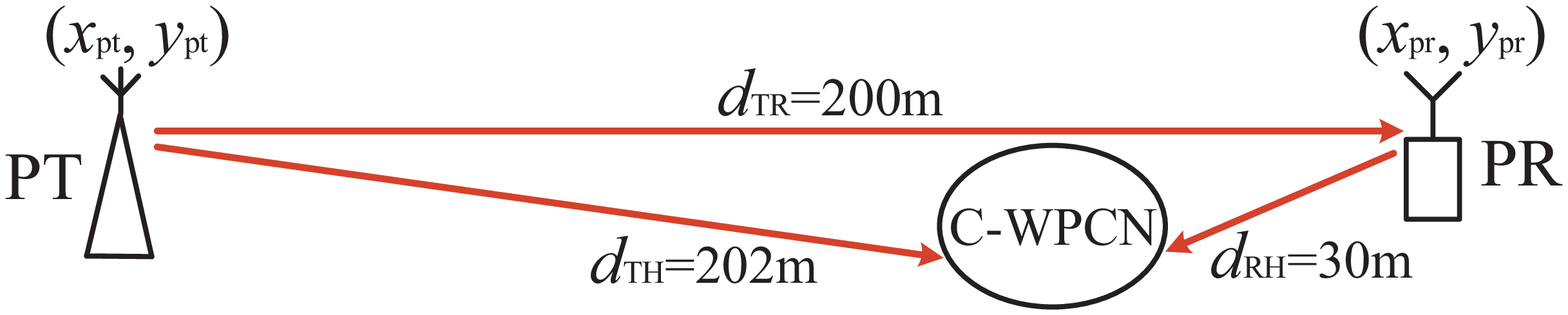}}
\subfigure[]{
\label{fig:subfig:b} 
\includegraphics[width=3.4in]{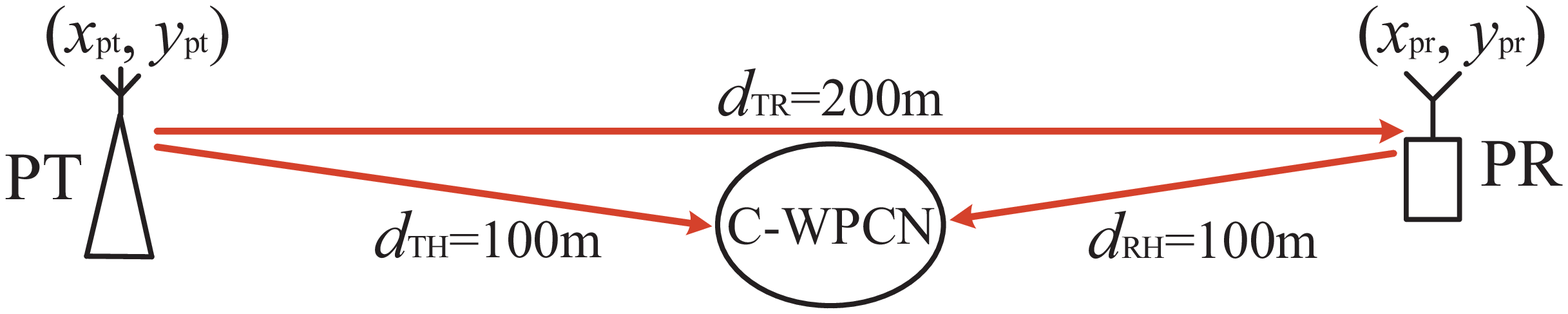}}
\subfigure[]{
\label{fig:subfig:c} 
\includegraphics[width=3.4in]{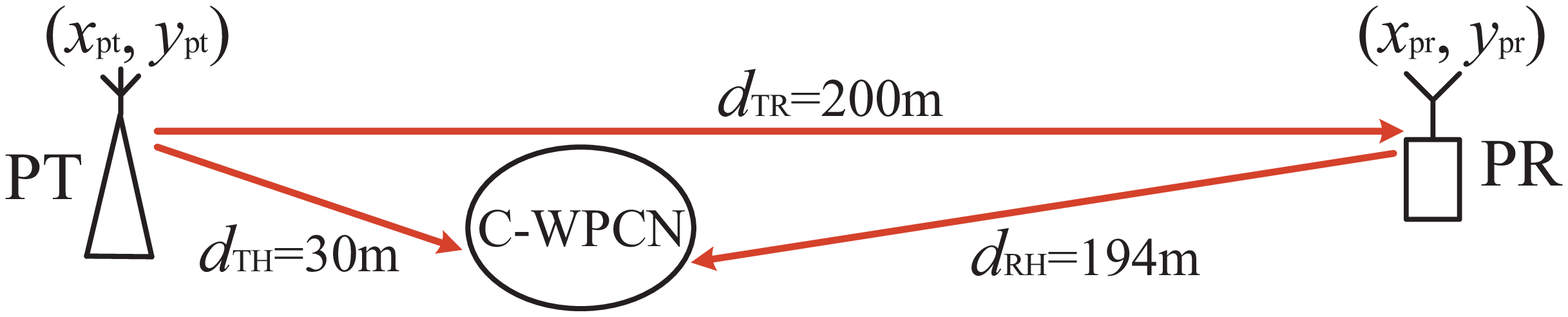}}
\caption{Simulation setup. (a) Case 1. (b) Case 2. (c) Case 3.}\Large
\label{107} 
\end{figure}
\subsubsection{Hybrid transmission}
In a hybrid transmission scheme, we separate the WDs into two groups. One group of WDs perform the cluster-based cooperation as proposed in this paper, while the WDs in the other group transmit independently to the HAP. In particular, we first determine the CH as the WD that is closest to the cluster center. Then, a WD chooses to transmit directly to the HAP if its channel to the HAP is better than that to the CH, otherwise, it operates under the previously proposed cooperation method. Specifically, the two groups transmit in orthogonal time to avoid interference. The transmission time and power of each WD, either transmitting independently to the HAP or cooperatively via the CH, is jointly optimized with those of the HAP. The detailed expressions are omitted here due to the page limit.
\begin{figure}
\centering
\subfigure[]{
\label{figa} 
\includegraphics[width=3.4in]{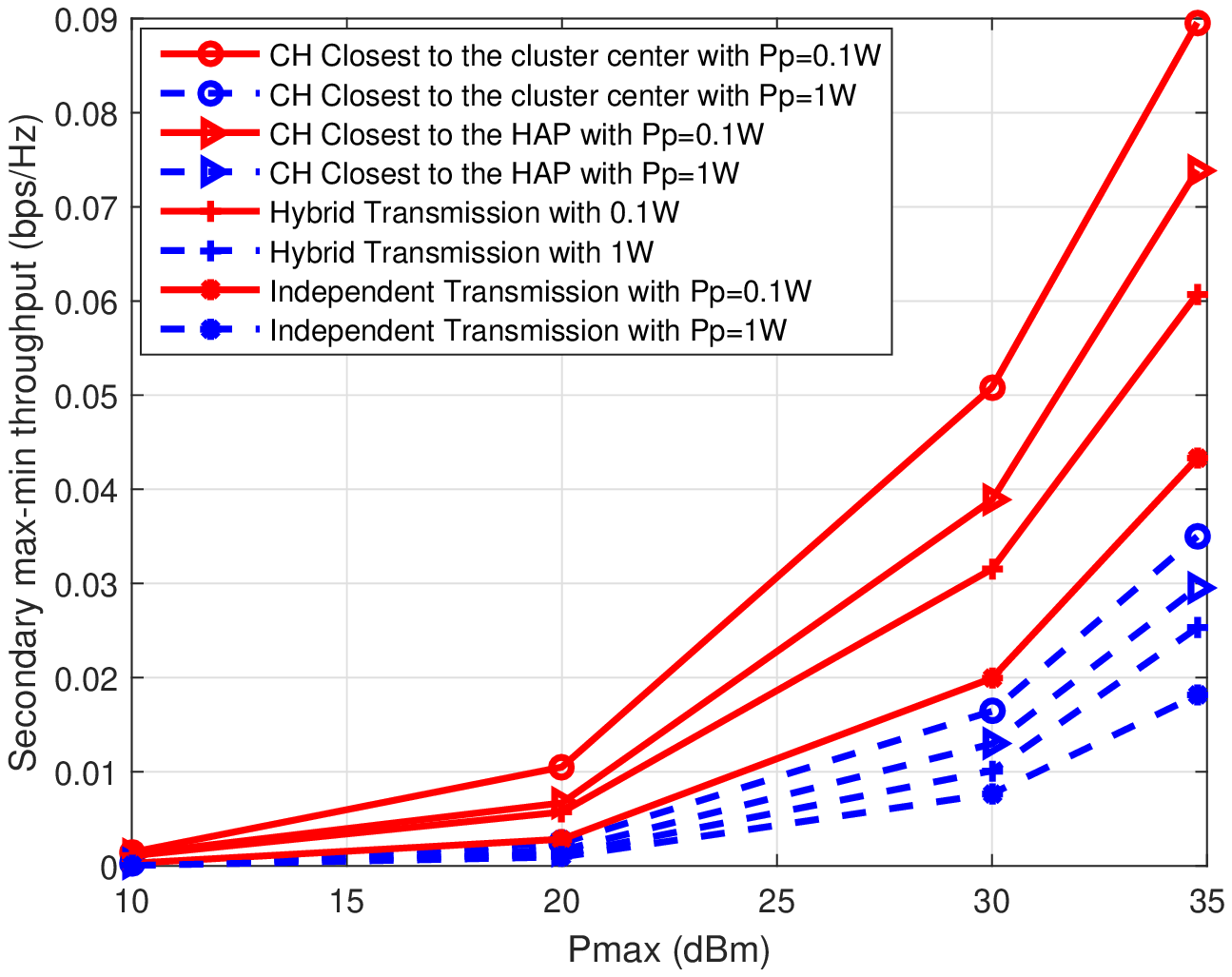}}
\subfigure[]{
\label{fig:subfig:b} 
\includegraphics[width=3.4in]{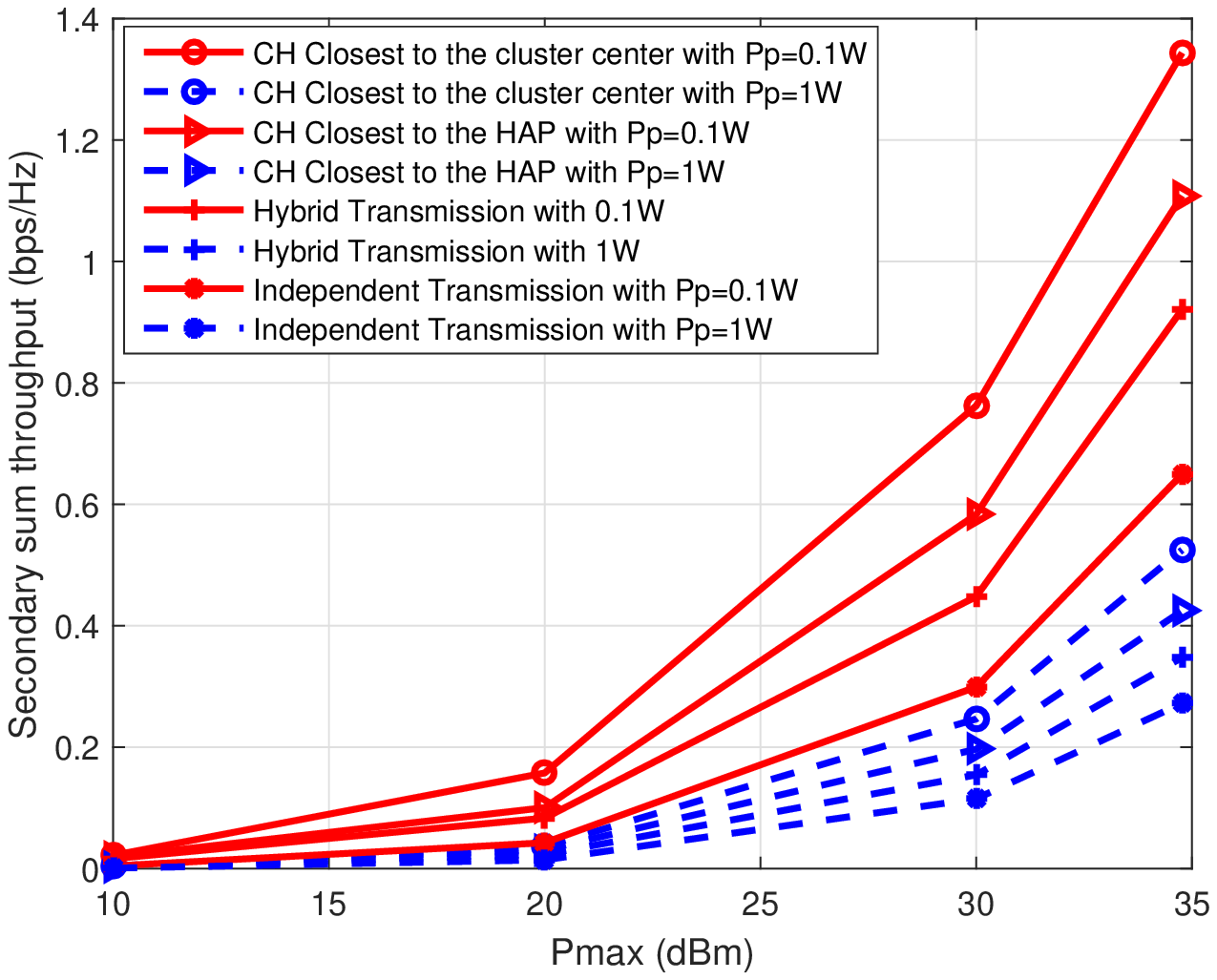}}
\caption{The maxi-min and sum throughput performance versus the maximum HAP transmit power $P_{max}$ for Case 1. (a) Max-min throughput performance. (b) Sum throughput performance.}\Large
\label{108} 
\end{figure}

\begin{figure}
\centering
\subfigure[]{
\label{figa} 
\includegraphics[width=3.4in]{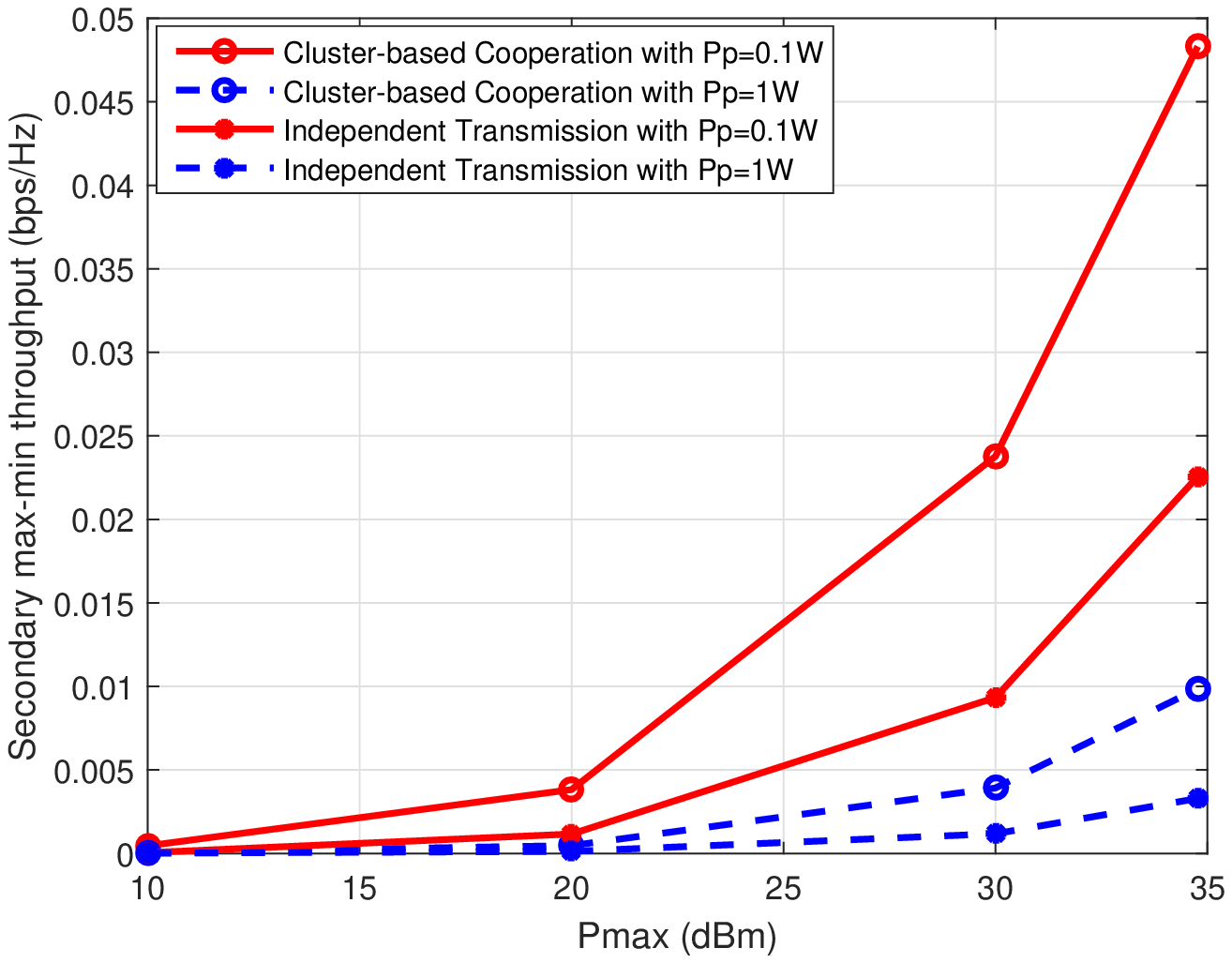}}
\subfigure[]{
\label{fig:subfig:b} 
\includegraphics[width=3.4in]{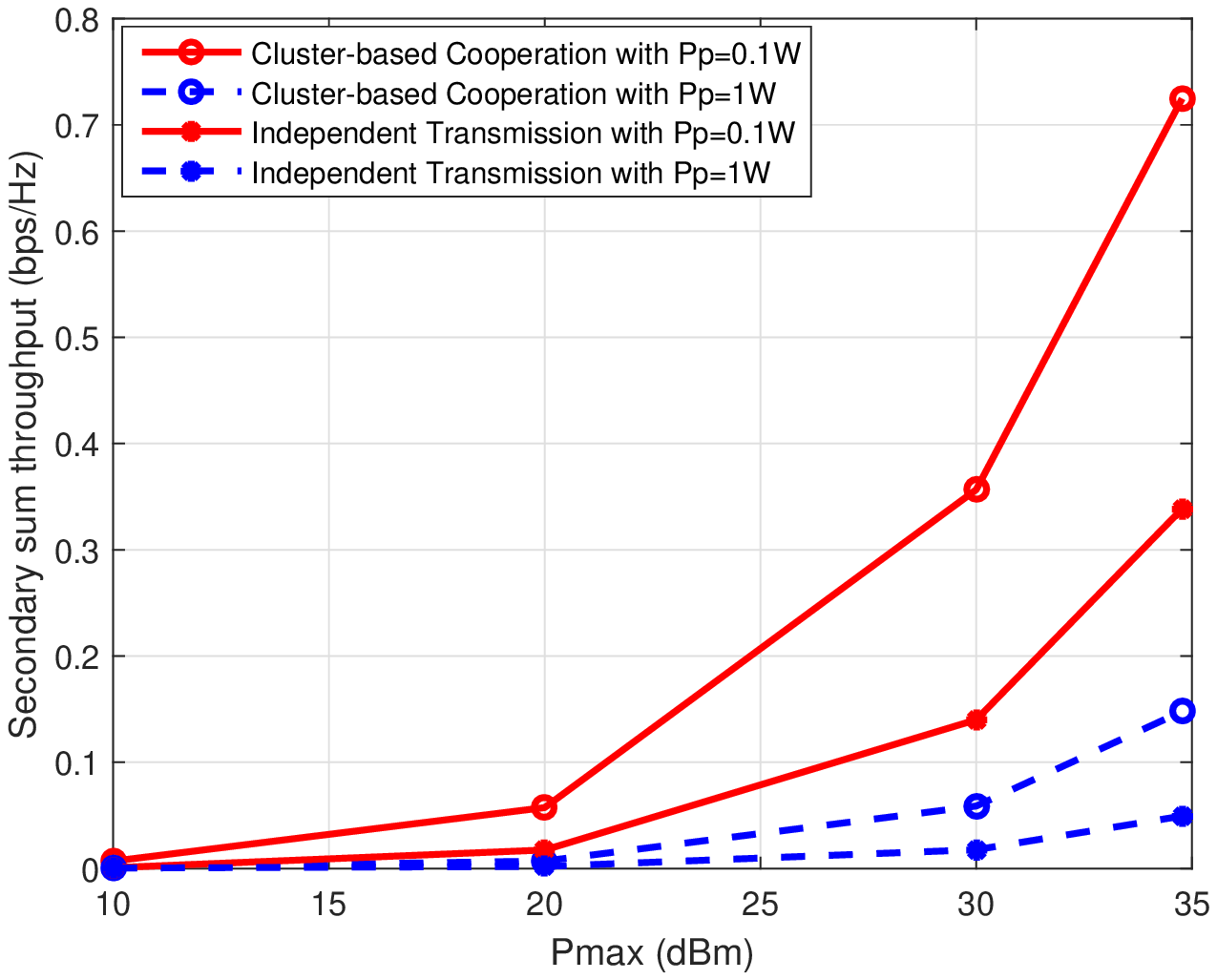}}
\caption{The maxi-min and sum throughput performance versus the maximum HAP transmit power $P_{max}$ for Case 2. (a) Max-min throughput performance. (b) Sum throughput performance.}\Large
\label{109} 
\end{figure}

\section{Simulation Results}
In this section, we evaluate the performance of the proposed cooperation method. In all simulations, we set the noise power as $N_{0} = 10^{-12}$ W and the receiver energy harvesting efficiency as $\eta= 0.5$. Unless otherwise stated, it is assumed that the number of antennas at HAP is $M=5$ and the threshold of the peak interference temperature constraint is $I_{max} = -60$ dBm (three orders of magnitude larger than the noise power, and in fact, later we discuss the performance of the system with different thresholds as shown in Figure 8.) in the considered bandwidth $P_p=0.1$ W and $P_{max}=3$ W. The mean channel gain between any two nodes (HAP, WD, PT and PR) follows a path-loss model. Without loss of generality, we assume the circuit energy consumption of all the devices are zero, i.e., $e_i = 0$, $i=0,\cdots,N-1$. For instance, let $d_{H,i}$ denote the distance between the HAP and the $i$-th WD, then the average channel gain $\delta_{H,i}^2 = G_A(\frac{3\times 10^8}{4\pi d_{H,i}f_c})^{\alpha}$, where $G_A$ denotes the antenna gain, $\alpha$ denotes the path-loss factor and $f_{c}$ denotes the carrier frequency. Likewise, $d_{PT,i}$, $d_{PR,i}$, $d_{CH,i}$, $d_{PR,H}$, $d_{PT,H}$, and $d_{PT,PR}$ denote the distance between the PT and the $i$-th WD, the PR and the $i$-th WD, CH and the $i$-th CM, the PR and the HAP, PT and HAP, and PT and PR, respectively, and their corresponding average channel gain model is similar to the above. We set $G_A=4$, $\alpha =3$, and $f_{c}=915$ MHz. Besides, $15$ WDs are uniformly distributed within a circle with radius equal to $r=3$ meters, and the circle's center is $d=6$ meters away from the HAP as shown in Fig. 3. Each point in the figures of all the simulations is an average of $100$ independent WD placements, while the performance of each placement is averaged over 1000 Rayleigh fading realizations \cite{2016:Bi2}. For simplicity, we assume the coordinates of the circle's center is $(0,0)$. In all simulations, the WD that is closest to the cluster center is selected as the cluster head.

The placements of the primary and secondary systems considered in this paper are shown in Fig. 4. We assume that the PT and PR are located 200 meters (m) apart. Meanwhile, we consider three different representative locations of the secondary WPCN. In Case 1, the C-WPCN is located much closer to the PR than PT (the HAP's coordinate of the C-WPCN is 202 m from the PT) as shown in Fig. 4(a); in Case 2, the distance from PT to C-WPCN and from PR to C-WPCN is equal (i.e., 100 m) as shown in Fig. 4(b); while in Case 3, the C-WPCN is located much closer to the PT (with a distance of 30 m from the PT) as shown in Fig. 4(c).
\begin{figure}
\centering
\subfigure[]{
\label{figa} 
\includegraphics[width=3.4in]{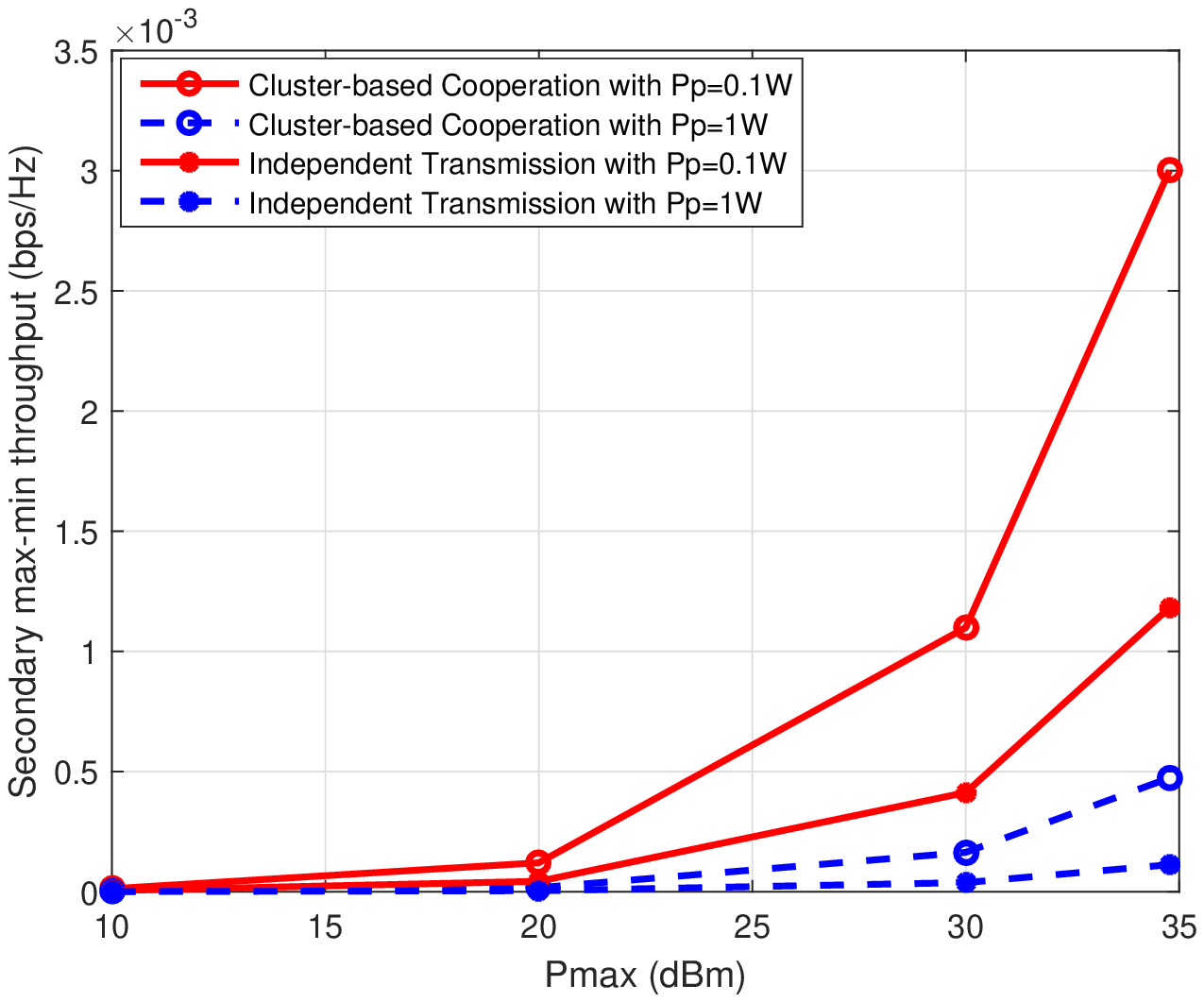}}
\hspace{1in}
\subfigure[]{
\label{fig:subfig:b} 
\includegraphics[width=3.4in]{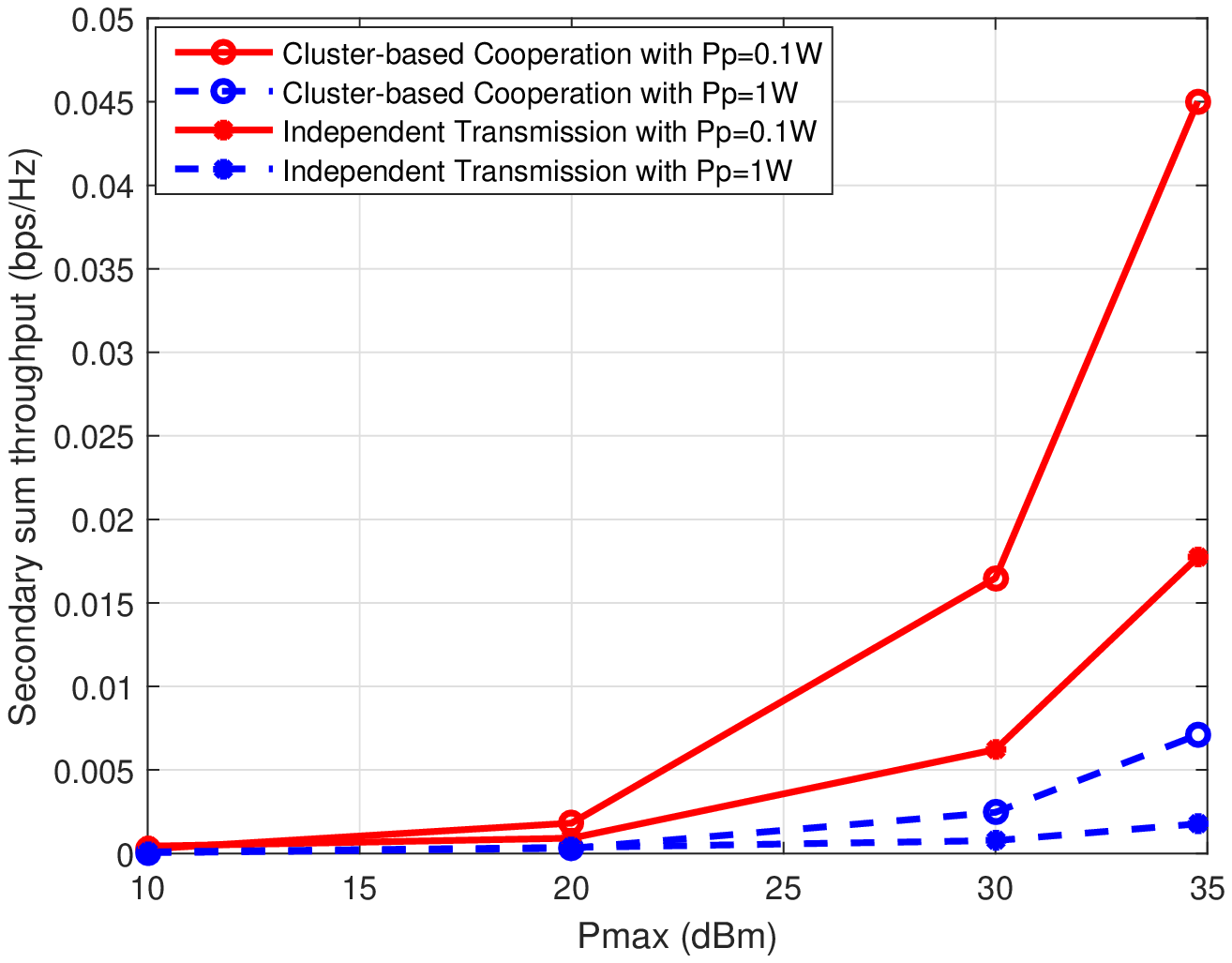}}
\caption{The maxi-min and sum throughput performance versus the maximum HAP transmit power $P_{max}$ for Case 3. (a) Max-min throughput performance. (b) Sum throughput performance.}\Large
\label{110} 
\end{figure}

\begin{figure}
\centering
\subfigure[]{
\label{figa} 
\includegraphics[width=3.4in]{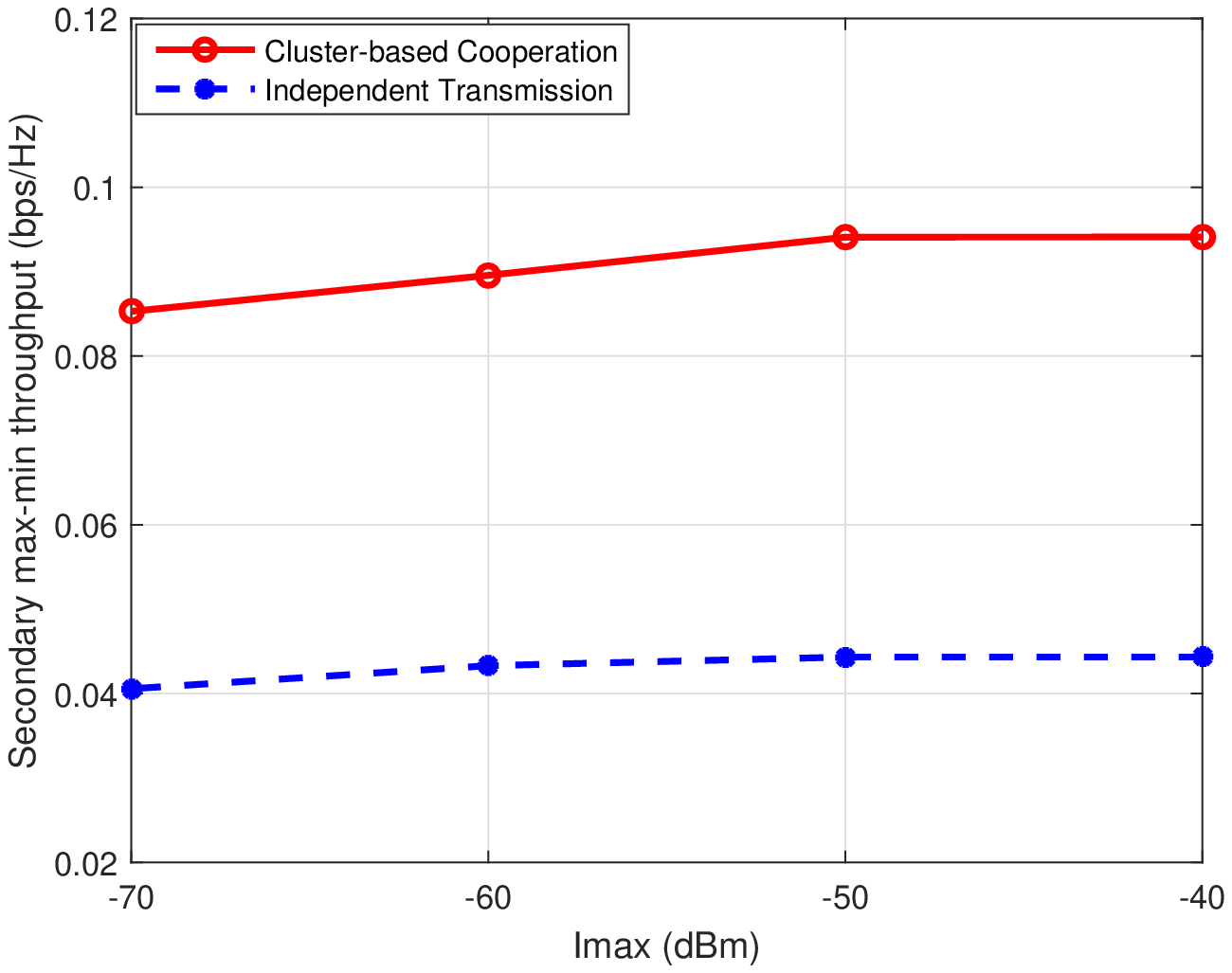}}
\hspace{1in}
\subfigure[]{
\label{fig:subfig:b} 
\includegraphics[width=3.4in]{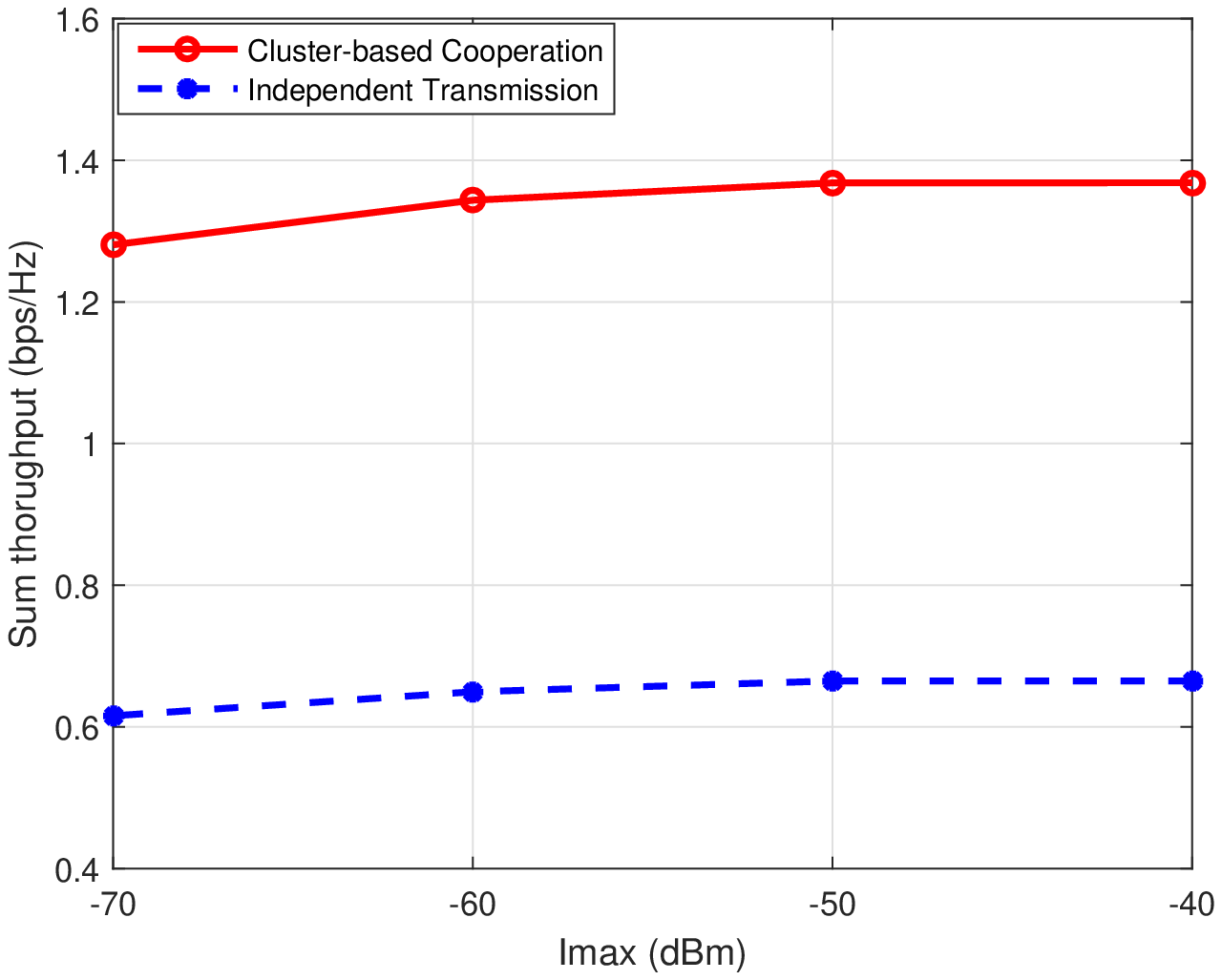}}
\caption{Performance comparison of the different transmission schemes when $I_{max}$ varies in Case 1. The figures above and below compare the max-min throughput and sum throughput, respectively.}\Large
\label{111} 
\end{figure}

\begin{figure}
\centering
\subfigure[]{
\label{figa} 
\includegraphics[width=3.4in]{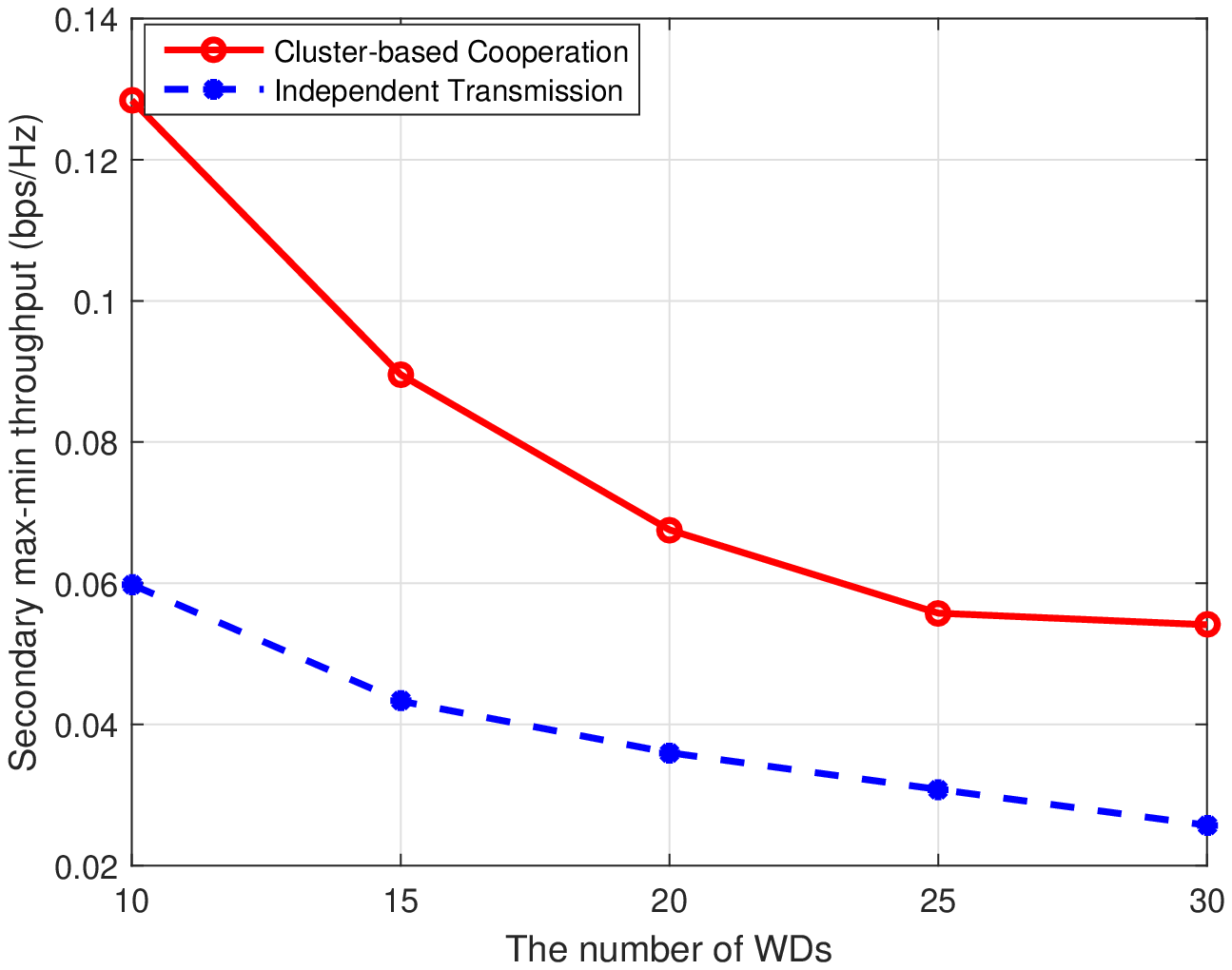}}
\hspace{1in}
\subfigure[]{
\label{fig:subfig:b} 
\includegraphics[width=3.4in]{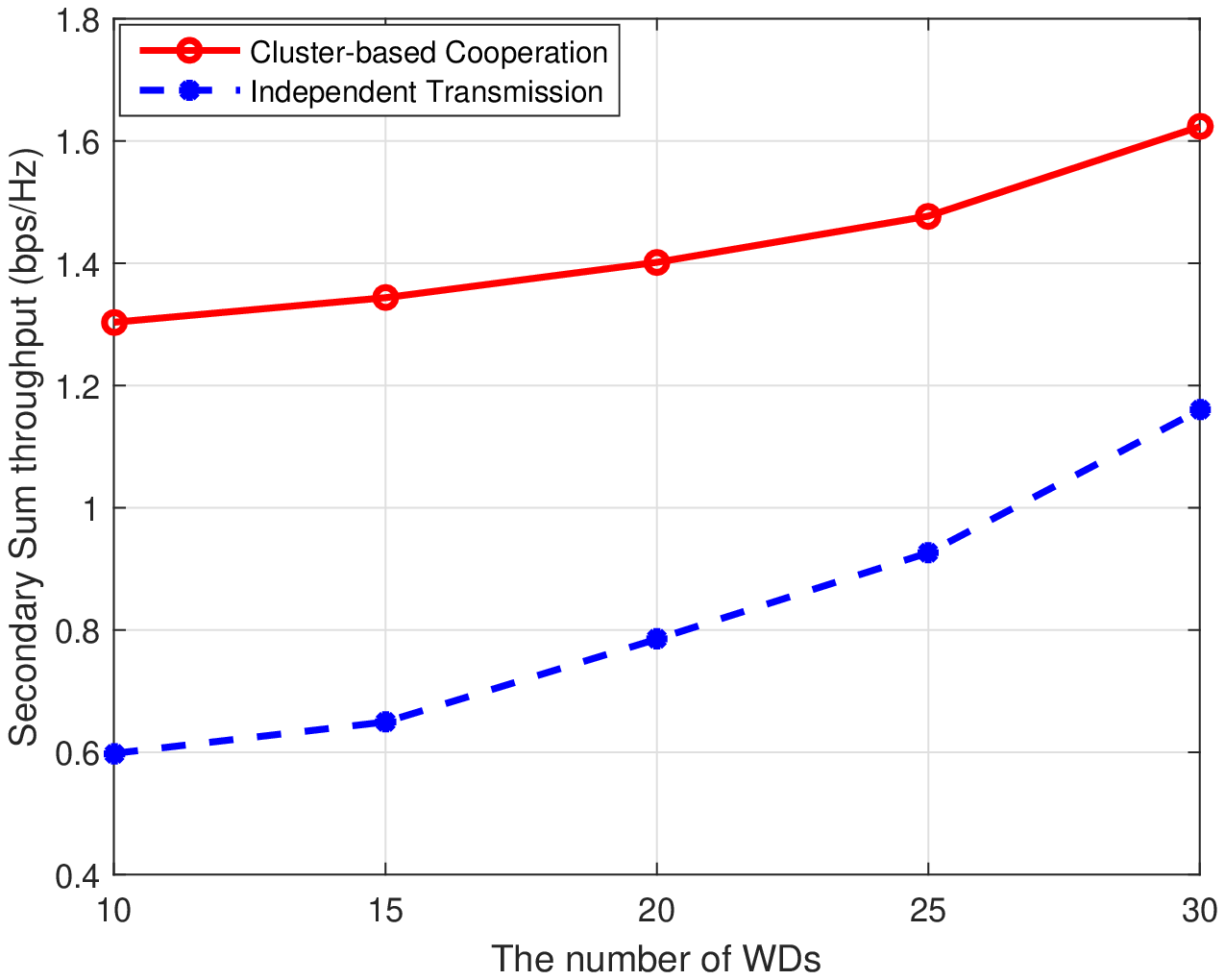}}
\caption{Performance comparison of the different transmission schemes when the number of WDs $N$ varies in Case 1. The figures above and below compare the max-min throughput and sum throughput, respectively.}\Large
\label{112} 
\end{figure}

In Fig. 5, we compare the max-min throughput performance of four methods, i.e., the proposed cluster-based cooperation (CC) selecting the CH as the WD closest to the cluster center, the CC selecting the CH as the WD closest to the HAP, the hybrid transmission, and independent transmission (IT) when the maximum transmit power of the HAP ($P_{max}$) varies. Meanwhile, we also consider two different cases where the transmit power of the PT $P_p=0.1$ W and $P_p=1$ W, respectively. The max-min throughput performance under the placement of Case 1 is shown in Fig. 5(a). We see that under both $P_p=0.1$ and $1$ W, the performance of the proposed CC method with CH closest to the cluster center is the best. Particularly, with $P_p=0.1$ W and $P_p=1$ W, it achieve on average around 26\% and 21\% higher max-min throughput than those of the CC with CH closest to the HAP, around 35\% and 32\% higher max-min throughput than those of the hybrid transmission, and around 56\% and 50\% higher max-min throughput than those of IT, respectively. Furthermore, Fig. 5(b) shows the sum throughput comparison under different $P_p$ in Case 1. It is observed that the proposed CC with CH closest to the cluster center performs the best. Specifically, it has on average 26\% and 23\% higher sum throughput than CC with CH closest to the HAP, around 35\% and 32\% higher sum throughput than those of the hybrid transmission, and average 56\% and 50\% higher sum throughput than IT with $P_p=0.1$ W and $P_p=1$ W, respectively. For simplicity of illustration, in the following simulations, we compare only the performance of the proposed CC method with CH closest to the cluster center (simply referred to as the CC method in the following) and the IT method, which is one commonly considered multi-user transmission method in related literatures.

In Fig. 6 and 7, we investigate the secondary max-min and sum throughput of CC and IT when $P_{max}$ changes for Case 2 and Case 3, respectively. Due to the closer distance between the primary and secondary systems, the secondary system suffers stronger interference from the primary network. Therefore, the data rates of the secondary network in Case 2 and 3 is worse than that in Case 1. In Case 3 when the two networks are only 30 meters apart, the achievable max-min throughput is only around 1/30 of that in Case 1. However, the proposed CC method still significantly outperforms the IT method in all simulation setups. Specifically, in Fig. 6(a), for Case 2 where the WPCN is equally distant from the PT and PR, the proposed CC method achieves on average around 56\% higher max-min throughput than the IT method when $P_p=0.1$ W and 58\% higher when $P_p = 1$ W. In terms of sum throughput performance in Fig. 6(b), the performance advantage of the proposed CC method is around 56\% and 58\% when $P_p=0.1$ W and 1 W, respectively. In Fig. 7, when the WPCN is closely located to the PT as in Case 3, the proposed CC method achieves on average around 61\% higher max-min throughput than the IT method when $P_p=0.1$ W and 68\% higher when $P_p = 1$ W. The sum throughput performance advantage is on average around 60\% and 68\% when $P_p=0.1$ and $1$ W, respectively. Overall, we can observe from Figs. 5-7 that the proposed CC method can achieve both higher user fairness and spectral efficiency in all the three representative network placements.

In Fig. 8, we plot the max-min and sum throughput performance of the proposed CC and IT when the threshold of the peak interference temperature constraint $I_{max}$ varies. For simplicity of illustration, we consider only the network setups in Case 1, while similar performance comparisons are also observed for Case 2 and 3. Here we consider the number of WDs $N=15$. We can see from Fig. 8 that the maxi-min and sum throughput of both the proposed CC and IT increase with $I_{max}$, i.e., higher tolerance of interference. Besides, both the max-min throughput and the sum throughput performance of the proposed CC method greatly outperform those of the IT method. Specifically, in Fig. 8(a), the proposed CC method achieves on average around one time higher max-min throughput than that of IT. Furthermore, in Fig. 8(b), the proposed CC has on average one time higher sum throughput than IT. Meanwhile, we see that the cluster-based C-WPCN is most effective under stringent ITC requirement, i.e., low $I_{max}$. For example, the performance advantage is more than one time for max-min throughput and one time for sum throughput when $I_{max}= -70$ dBm. Intuitively, this is because the cluster-based cooperation can effectively control its interference to the primary system by reducing the communication range, and thus the transmit power, when the secondary users transmit their information.

In addition, Fig. 9 evaluates the throughput performance when the number of WDs $N$ increases from 15 to 30 in Case 1. On one hand, we can see from Fig. 9(a) that the max-min throughput decreases with the number of WDs for both schemes. This is because in general each WD is allocated with shorter transmission time as $N$ increases, especially for the time allocated to the worst-performing WD due to the doubly-near-far unfairness issue. In an extreme case, when the number of the WDs approaches infinity, the transmission time allocated to the worst-performing WD approaches zero, thus resulting close-to-zero max-min throughput. Nonetheless, we see that the proposed CC method achieves around 100\% higher max-min throughput than the benchmark IT method. On the other hand, we can observe in Fig. 9(b) that the sum-throughput increases with $N$ due to the benefit of multi-user diversity, although the data rates of some individual WDs may decrease. In particular, the proposed CC method achieves on average around one time higher sum throughput than the IT method. This indicates that a tradeoff exists between each individual user's throughput and the aggregate network throughput. In practice, the number of WDs should be kept moderate, to guarantee satisfying per-user throughput performance. Nonetheless, we can still observe significant performance gain of the proposed method over the benchmark method.

\vspace{-2ex}

\section{Conclusions}
In this paper, we have considered a cluster-based cooperation method in a WPCN underlaid to a primary communication link. Specifically, a CH user is appointed to relay the information transmission of the other cluster members to the HAP. Energy beamforming is applied at the multi-antenna HAP to balance the different energy consumption rates of the WDs. To control the potential severe interference, the secondary WPCN generates to the primary system, we optimize the system performance under an interference temperature constraint. By jointly optimizing the energy beamforming design, the transmit time allocation among the HAP and the WDs, and the transmit power allocation of the CH. Extensive simulations under practical network setups show that the proposed method can significantly enhance user fairness and spectrum efficiency compared to the benchmark method, meanwhile guaranteeing the quality of transmission in the primary network, especially under stringent ITC requirement.

\end{document}